\documentclass[twocolumn]{aastex63}

\usepackage{longtable}
\usepackage{lipsum}

\accepted{\today}

\submitjournal{ApjL}

\shorttitle{Kepler-411 activity: starspots versus superflares}
\shortauthors{Araújo \& Valio}

\graphicspath{{./}{figures/}}

\begin{document}

\title{Kepler-411 star activity: connection between starspots and superflares}

\correspondingauthor{Alexandre Araújo}
\email{adesouza.astro@gmail.com, avalio@craam.mackenzie.br}

\author[0000-0002-2106-4332]{Alexandre Araújo}

\author[0000-0002-1671-8370]{Adriana Valio}

\affiliation{Center for Radio Astronomy and Astrophysics Mackenzie, Mackenzie Presbyterian University\\
Rua da Consolação, 860, São Paulo, SP - Brazil}

\begin{abstract}
Stellar magnetic activity, just like that of the Sun, manifests itself in the form of flares and spots on the surface of the star. In the solar case, the largest flares originate from large active regions.
In this work, we present a study of the activity of the star Kepler-411, including spot modeling from planetary transits. Our goal was to search for a connection between the area of starspots with the energy of superflares produced by this star. Kepler-411 is a K2V-type star with an average rotation period of 10.52 days, radius of 0.79 $R_{\odot}$ and a mass of 0.83 $M_{\odot} $, which was observed by the Kepler satellite for about 600 days. Transit mapping allowed for the characterization of 198 starspots with estimates of their radius and temperature. Kepler-411 starspots had an average radius of $(17 \pm 7)\times 10^3$ km and a mean temperature of $3800 \pm 700$ K.
Visual inspection of the light curves of  Kepler-411 yields the identification of 65 superflares. The detected superflares lasted from 8 to 260 min and their energy varied from $10^{33} - 10^{35}$ ergs. 
The power-law index of the flare 
frequency distribution as a function of energy is (-2.04 $\pm$ 0.13) for the flare on Kepler-411.
A positive correlation between the area of starspots and the energy of superflares was found when considering the averages taken every 16 to 35 days, with the highest correlation occurring for  averages every 21 days. This timing is probably related to the lifetime of the Kepler-411 spots. 

\end{abstract}


\section{Introduction} \label{sec:intro}

Astrophysics has benefited enormously from observations of space missions such as CoRoT (COnvection ROtation et Transits planétaires), Kepler, and TESS (Transiting Exoplanet Survey Satellite). The data collected from these missions are crucial for understanding the formation of our Solar System, the relation of the star and its planets, planetary habitability, the emergence of life, and the mechanisms involving the formation of the various exoplanets.\\

Data from the Kepler telescope are an excellent resource for research on stellar activity. This data set, involving unprecedented photometric precision and continuity over four years, resulted in the discovery of a vast diversity of exoplanets and made it possible to study several phenomena related to stellar activity.\\

Several models were applied to complete light curves to study starspots, however the techniques still present problems as reported in \cite{Basri2020}. Alternatively, the presence of spots on the stellar surface can be detected by analyzing the planetary transits light curve of the star
\citep{silva+03}. The passage of the planet in front of the star can reveal spots on the star's surface. \cite{silva+03} proposed the first  model to use the transit of an exoplanet as a probe to detect the presence of spots on the surface of stars, physically characterizing spots of solar-type stars (F, G, K) and M dwarfs. 
Various physical properties and position information about stellar spots have been successfully estimated for solar-type  and M dwarf stars from transit mapping \citep{silva2010, SanchisOjeda2013,  valio+17, morris2017starspots, zaleski2019, zaleski2020, netto2020,  namekata2020temporal}. In a recent study, using transit mapping it was possible to estimate the differential rotation of Kepler-411 from the transit of its three orbiting exoplanets \citep{araujo+21}.\\

Several  studies about flares and stellar superflares have provided an increasing wealth of information for stars of different spectral types. 
Spectroscopic studies highlight the idea that the variation in brightness of superflare stars is correlated with the rotation of stars with large  spots \citep{notsu2015high, notsu2019kepler}. This discovery points to a relation between superflares and starspots, and hence magnetic fields. Solar flares usually occur in association with sunspots or groups of spots in active regions.\\

Sunspots correspond to regions of more intense magnetic fields than the rest of the Sun's surface, being colder and darker than the solar disk \citep{hale}. Like the spots on the Sun, other stars have these phenomena which are related to star activity. Probably, all cool stars with a convective region similar to that of the Sun  have spots on their surface, according to \cite{strassmeier}.\\

\cite{karoff2016observational} observed indications of a correlation between the S-index and the periodic photometric variability amplitude down to amplitudes around 1,000 ppm, which is the level of amplitudes observed for the Sun. Below 1,000 ppm, no clear indication of a correlation was seen, indicating that the relation between spot coverage and chromospheric emission breaks down when  magnetic activity is low.\\ 
 
Kepler space high precision photometry \citep{borucki2010kepler} with its long duration
enabled the systematic study of stellar flares, improving our understanding of stellar activity. For example, \citet{walkowicz+11} studied about 23 thousand dwarf stars and identified 373 stellar flares. In this study, the authors calculated the relative energies and occurrence rate of flares, and found that M dwarfs tend to produce flares more frequently and with shorter duration than K dwarfs.\\

Studies of flares and superflares have been extended to solar-type stars. \cite{maehara+12,maehara+15} searched for superflares in solar-type stars using  long and short cadence Kepler data, and found that all observed superflares occur in stars with large starspots, evidenced by the almost periodic modulation of the star brightness.
Also, 187 superflares were detected in 23 solar-type stars which bolometric energy ranging from $10^{32}$ to $10^{36}$ erg. These superflares showed a complex temporal behavior with several peaks of a separation of the order of 100 to 1000s, comparable to the periods of quasi-periodic pulsations in solar and stellar flares.\\ 

\cite{shiba+13} studied superflares in solar-type stars using Kepler data for a longer time period (500 days). As a result, 1547 superflares were found in 279 stars. In a recent publication, \cite{okamoto2021statistical} updated the statistical information for solar-type stars (G) using all data from the primary Kepler mission and the Gaia Data Release 2 catalog. With the new results, a total of  2341 superflares were found in 265 solar-type stars and 26 superflares in 15 stars similar to the Sun.\\ 

Moreover,  it is known that the frequency of superflares decreases as the stellar rotation period increases, or as the star ages. The maximum flare energy  found on Sun-like stars is $4 \times 10^{34}$ erg, and analysis of Sun-like stars suggests that the Sun can cause superflares with energies up to $\sim7 \times 10^{33}$ erg (equivalent to X700-class flares) and $\sim 1 \times 10^{34}$ erg (X1000-class flares) once every $\sim$3000 and $\sim$6000 yr, respectively \citep{okamoto2021statistical}.\\

Investigations about stellar activity have also been the subject of studies using Kepler data, focusing on the relation between flares and starspots. The work of \cite{roett+18} investigated a subset of Kepler's observations and revealed a connection between starspots and stellar flare. The study provided valuable information about the star magnetic field,  by emphasizing that many of the stars observed by Kepler behave significantly different from the Sun, with larger starspots. \cite{hawlay+14} also investigated the correlation with starspot modulation. The results of the starpot phase vs energy ratio of flares show that the flare energy is randomly distributed and does not show strong correlation with the starspot phase. Similar results were also found by  \cite{maehara2021time}, \cite{silverberg}, and \cite{doyle2020superflares}, that is, flares are randomly distributed.\\

Still challenging is the correlation of stellar flares and superflares with the presence of spots on the surface of the stars. Understanding the relation between flares and starspots is extremely important, not only for the dynamo mechanism at work on these stars, but also because these processes drastically affect the atmospheric and biological environment of orbiting planets. As a consequence, it can affect the habitability of these exoplanets. A broader approach to the impacts on the atmospheric environment of exoplanets can be seen in \cite{airapetian2020impact}.\\

For the first time, it is possible to study the connection of starspots with the presence of flares and superflares using the method of starspots mapping  from planetary transits. 
With information on the coverage of spots at different longitudes and latitudes, such as their sizes and frequency, we propose to search for correlations of these with the occurrence of flares and superflares from the star Kepler-411.
The study presented here of  Kepler-411  yields a more robust result with higher spatial precision of starspot sizes and location during each of the three planetary transits.\\

In the next section, we present the results of Kepler-411 activity, that is,  spots physical properties and flare identification, as well as estimates of their  energy and duration. In Section~\ref{sec:spotflares}, we compare the timing of spot area and flare energy searching for a correlation between them. Finally, in Section~\ref{sec:conclusao}, the results are discussed and the main conclusions presented.

\section{Kepler-411 activity: spots and superflares} \label{sec:method}

\begin{table*}
\setlength{\arrayrulewidth}{2\arrayrulewidth}
\centering
\caption{Physical parameters of Kepler-411 star, planets, and spots.}\label{tab:spot}
\def\arraystretch{1.2}  

\begin{tabular}{lcccc}
\hline
\hline
\multicolumn{5}{c}{\textbf{Stellar Parameters}}\\
\hline 
\hline
Spectral type  & Radius [$R_{\odot}$]$^{a}$ & T$_{eff}$[K]$^{b}$ & Period [days]$^{d}$ & Mass [$M_{\odot}$]$^{a}$\\

K2V  & $0.79^{+0.07}_{-0.06}$ & 4832 & $10.4 \pm 0.03$ & $0.83^{+0.04}_{-0.10}$\\
\hline
\hline
\multicolumn{5}{c}{\textbf{Planetary Parameters}}\\
\hline
\hline
 & Kepler-411b & Kepler-411c & Kepler-411d \\
\hline
Orbital Period [days]$^{a}$   & $3.0051 \pm 0.00005$ & $7.834435 \pm 0.000002$ & $58.02 \pm 0.0002$ & \\
Planet Radius [R$_\oplus$]$^b$ & $1.88 \pm 0.02$ & $3.27^{+0.011}_{-0.006}$ & $3.31 \pm 0.009$ & \\
Planet Radius [$R_{star}$]$^c$ &   $0.024 \pm 0.002$ &  $0.042  \pm 0.002$ &  $0.040 \pm 0.002$ \\
Semi-Major Axis [au]$^c$  & $0.049  \pm 0.0006$ &
  $0.080 \pm 0.001$   &  $0.29 \pm 0.0004$ & \\
Orbital inclination$^c$       & $89.18 \pm 0.05$ & 
     $89.03 \pm 0.02$ & $89.44 \pm 0.01$ & \\
\hline
\hline
\multicolumn{5}{c}{\textbf{Spot average parameters}}\\
\hline
\hline
Transits of & Kepler-411b & Kepler-411c & Kepler-411d & all \\
 \hline
Number of spots$^{c}$ & 45 & 143 & 10 & 198 \\
Temperature [K]$^{c}$ &  $3100\pm 800$ & $4100\pm 500$ & $3600\pm 800$ & $3800 \pm 700$\\
Radius [$10^3$ km]$^{c}$ & $18\pm 7$ &  
$17\pm 8$ & $17\pm 5$  & $17 \pm 7$ \\
Intensity [$I_c$]$^{c}$ & $0.14\pm 0.23$ & $0.46\pm 0.28$ & $0.29\pm 0.19$ & $0.35 \pm 0.24$\\
\hline
\end{tabular}
\tablecomments{{$^{a}$\cite{wang+14} $^{b}$ GAIA archive $^{c}$\citep{araujo+21} $^{d}$\citep{sun2019kepler}}.}
\end{table*}

\begin{table*}[ht]
\caption{Average parameters of superflares from Kepler-411.}
\label{tab:flare}
\begin{center}
\begin{tabular}{lc}
    \hline
    \hline
    \multicolumn{2}{c}{\textbf{Superflares average parameters}}\\
   \hline
   \hline
Decay time [Min] & $28 \pm5$\\ 
Rise time [Min] &  $5.1 \pm2.0$  \\
Duration [Min] &   $33 \pm6$ \\
Log energy [Erg] & $33.8 \pm0.3$\\
Flare frequency [Day$^{-1}$]    & 0.11             \\
\hline
\end{tabular}
\end{center}
\end{table*}

\begin{figure*}
 \centering
   \includegraphics[scale=0.5]{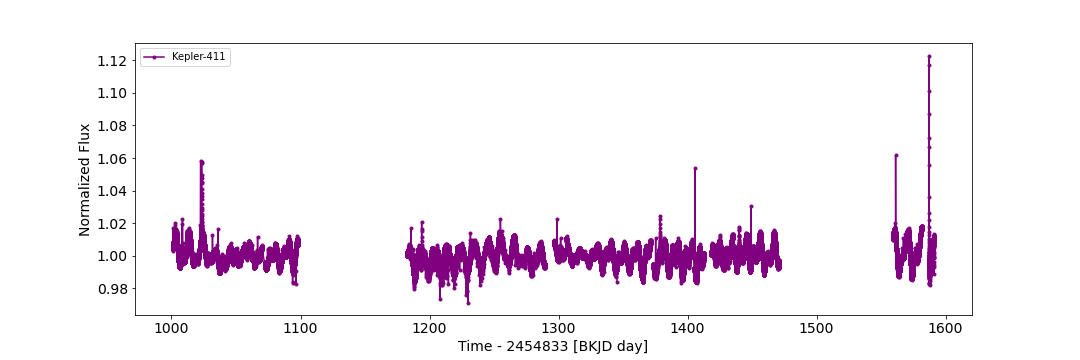}
    \includegraphics[scale=0.5]{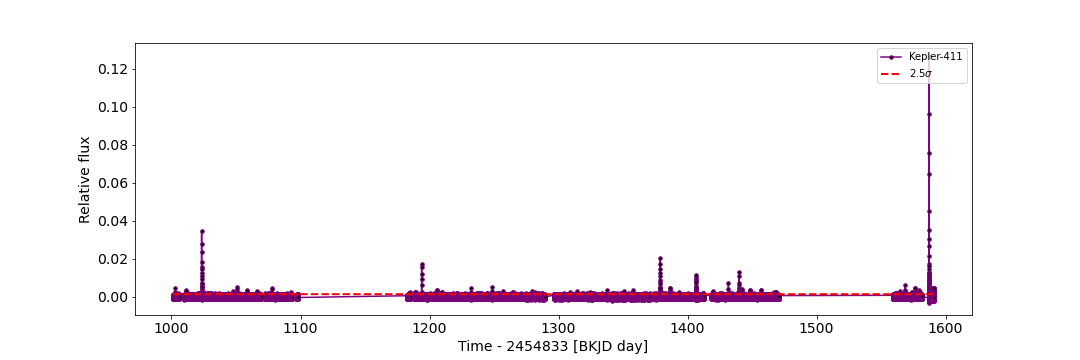}
   \caption{\textbf{Top:} Kepler-411 light curve normalized without corrections. \textbf{Bottom:} Light curve with corrections for analysis and identification of flares and superflares. }
    \label{fig:curva+flare}
\end{figure*}

Kepler-411 (KIC 11551692) was observed by the Kepler space telescope for about 600 days, exhibiting characteristics that indicate relatively strong magnetic activity \citep{sun2019kepler}, with many flares detected as can be seen in Figure~\ref{fig:curva+flare} (top panel). Kepler-411 is a star of spectral type K2V with three transiting planets, one SuperEarth and two mini-Neptunes.
Here we present a study based on the results obtained by \cite{araujo+21}, which mapped the spots at different latitudes, according to the various planetary transits. 

\subsection{Starspot modeling and physical parameters} \label{sec:starspotmodeling}

In the analysis of Kepler-411 light curves, we isolated the transits of the three planets, Kepler-411b, Kepler-411c, and Kepler-411d. There were 121, 48, and 7 planetary transits detected for Kepler-411b, Kepler-411c, and Kepler-411d exoplanets, respectively. As the planet crosses in front of the star during its transit, it may occult a spot, or dark region, on the surface of the star, causing small increases in the transit light curve. These small short duration increases due to starspots can be modeled and its physical characteristics determined  using the spot mapping model of \citet{silva+03}. Starspot detection is performed by visual inspection of each transit in the light curve. A model of a spotless star is subtracted from each transit light curve, so as to enhance spots signatures. Only peaks in the residual light curve with a size above 3$\sigma$ are considered as spots and modeled, never exceeding the maximum number of 4 spots per transit. Spots detected by visual inspection are then modeled and limited between longitudes of $\pm 70^\circ$ to avoid any distortions caused by the steep portions of the light curve due to the ingress and egress of planetary transit.\\ 

Each starspot is modeled as a circular disc with the following parameters:
\begin{itemize}
    \item Radius: size in units of the planetary radius ($R_p$);
    \item Intensity: photospheric contrast with respect to the maximum brightness of the star at disc center ($I_c$);
    \item Position: longitude and latitude on the stellar disc. 
\end{itemize}
The latitude, $lat$, and longitude, $lg$, of each spot are given by the projection of the planetary position onto the stellar disc.
Zero longitude was considered as the projection during the first midtransit of the planets, whereas the spot longitude, $lg$, is obtained from the time of the spot peak in the residuals, with $t_s$ given in hours, according to the following equations:
\begin{equation}
lat = \arcsin\left[\frac{a}{R_{star} \cos(i)}\right]
\end{equation}
\begin{equation}
    lg=\arcsin\left[\frac{a\ \cos [90^{\circ}-(360^{\circ}{t_s}/24~ P_{orb})]}{cos(lat)}\right],
     \label{eq:lg}
\end{equation}
\noindent where $R_{star}$ is the radius of the star; and the planet orbital parameters are: $a$  the semi-major axis, $P_{orb}$ the orbital period, and $i$ the orbital inclination angle with respect to the stellar rotation axis. Here, we assumed the stellar spin to have null obliquity.\\

The spot temperature is obtained assuming that both the star and the spot emit radiation like a black body (following Eq. 2 of \citet{silva2010}), 
where we consider $T_{eff} = 4832$K as the effective temperature of the stellar photosphere and an observing wavelength of $\lambda = 600$nm.
The statistical results of the spot parameters from Kepler-411 discussed in \cite{araujo+21} are listed in Table~\ref{tab:spot}.

\begin{figure}
    \centering
    \includegraphics[scale=0.21]{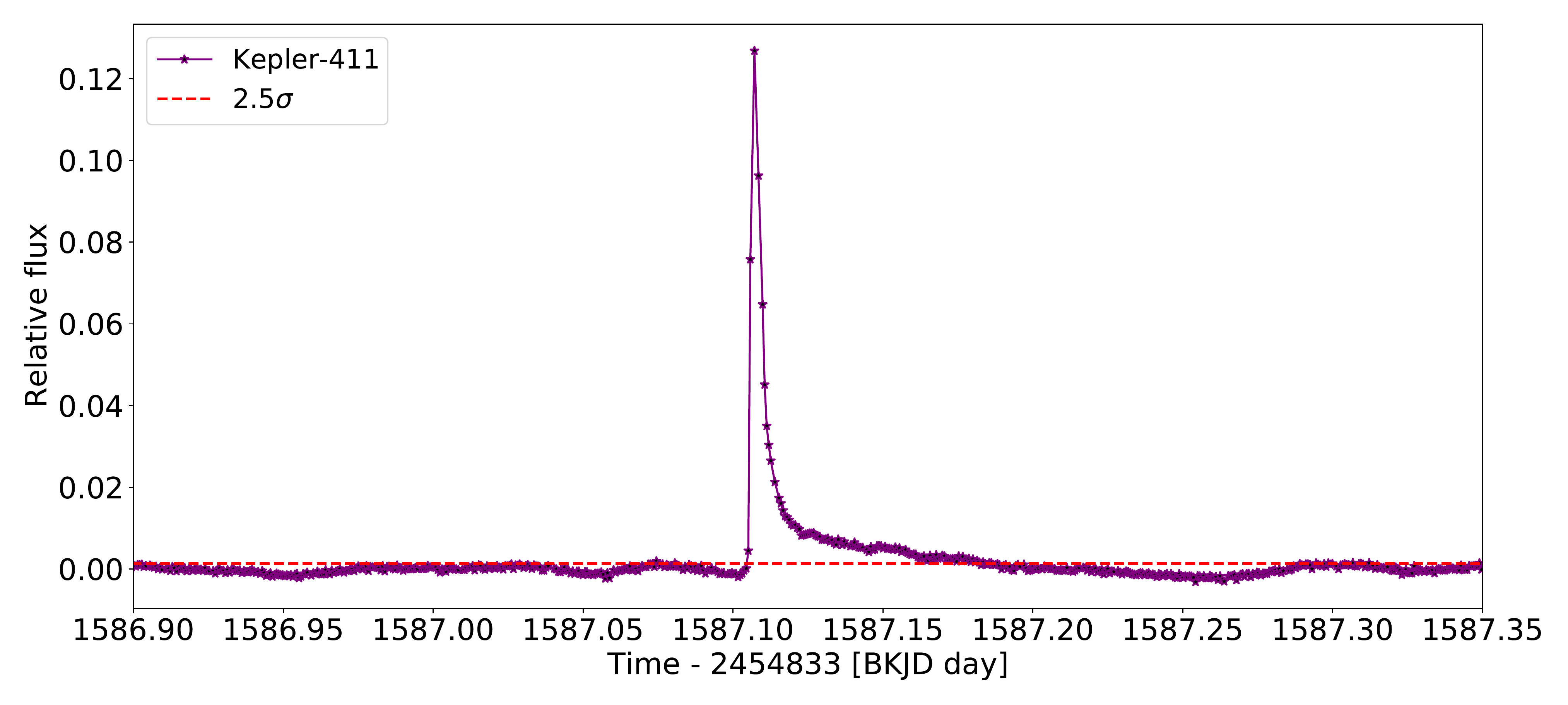}
    \caption{Largest superflare observed in the Kepler-411 light curve.}
    \label{fig:flarenacurva}
\end{figure}

\subsection{Flare identification and properties} \label{sec:flarecurva}
To demonstrate the steps used to identify and extract the superflares seen in the out--of--transit light curve of Kepler-411, we use as an example an excerpt from the light curve shown in Figure~\ref{fig:curva+flare}. In the upper panel is the light curve in PDC format for the entire period of observation for Kepler-411, quarters 11 – 17.
To remove the oscillatory trend due to the rotation of the spotted star from the light curve, a polynomial of degree three was applied and subtracted. The result is shown in the bottom panel of Figure~\ref{fig:curva+flare}, where also outliers and cosmic rays have been removed and the relative flux is obtained using Equation 2 of 
\cite{hawlay+14}.\\

Then, each quarter of the light curve was visually inspected and the superflare candidates  identified as three or more consecutive points with flux above the overall average flux by at least 2.5 standard deviations ($\sigma$). An example of a superflare detected this way is shown in Figure~\ref{fig:flarenacurva}, this is the largest superflare detected from Kepler-411 during Kepler observation.\\

Flare time profiles are normally divided into an impulsive phase followed by a gradual decay phase. It is important to note that the flares and superflares considered in this sample have a temporal profile similar to solar flares, that is  flares with a rise time shorter than the decay time. Flares with complex time profiles are described in \cite{kowalski2010white}. To avoid potential false positives, we used the classic definition of a flare  with light curve profiles such as that of Figure~\ref{fig:flarenacurva}.

\begin{figure}[!t]
\centering
\includegraphics[scale=0.22]{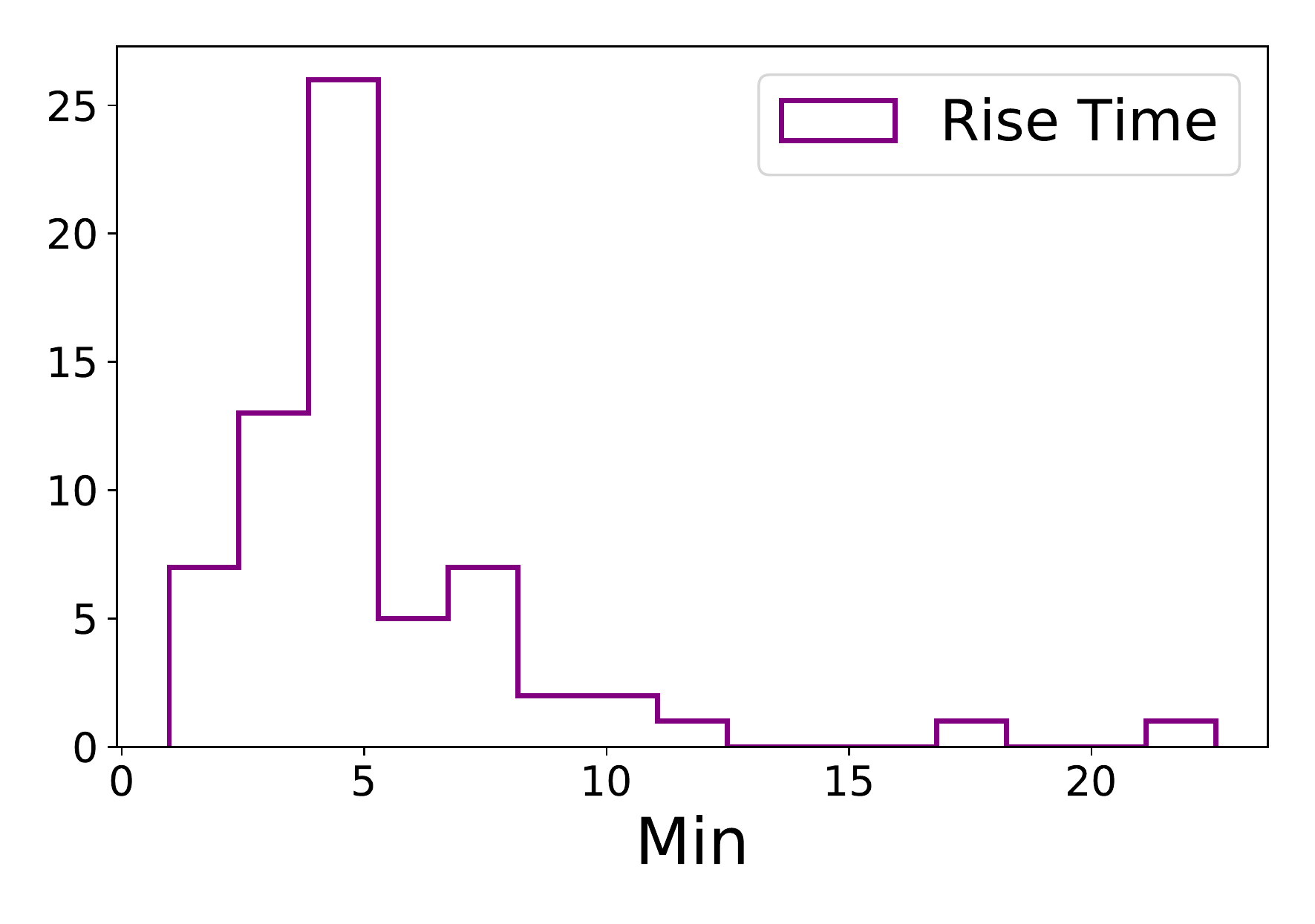}
\includegraphics[scale=0.22]{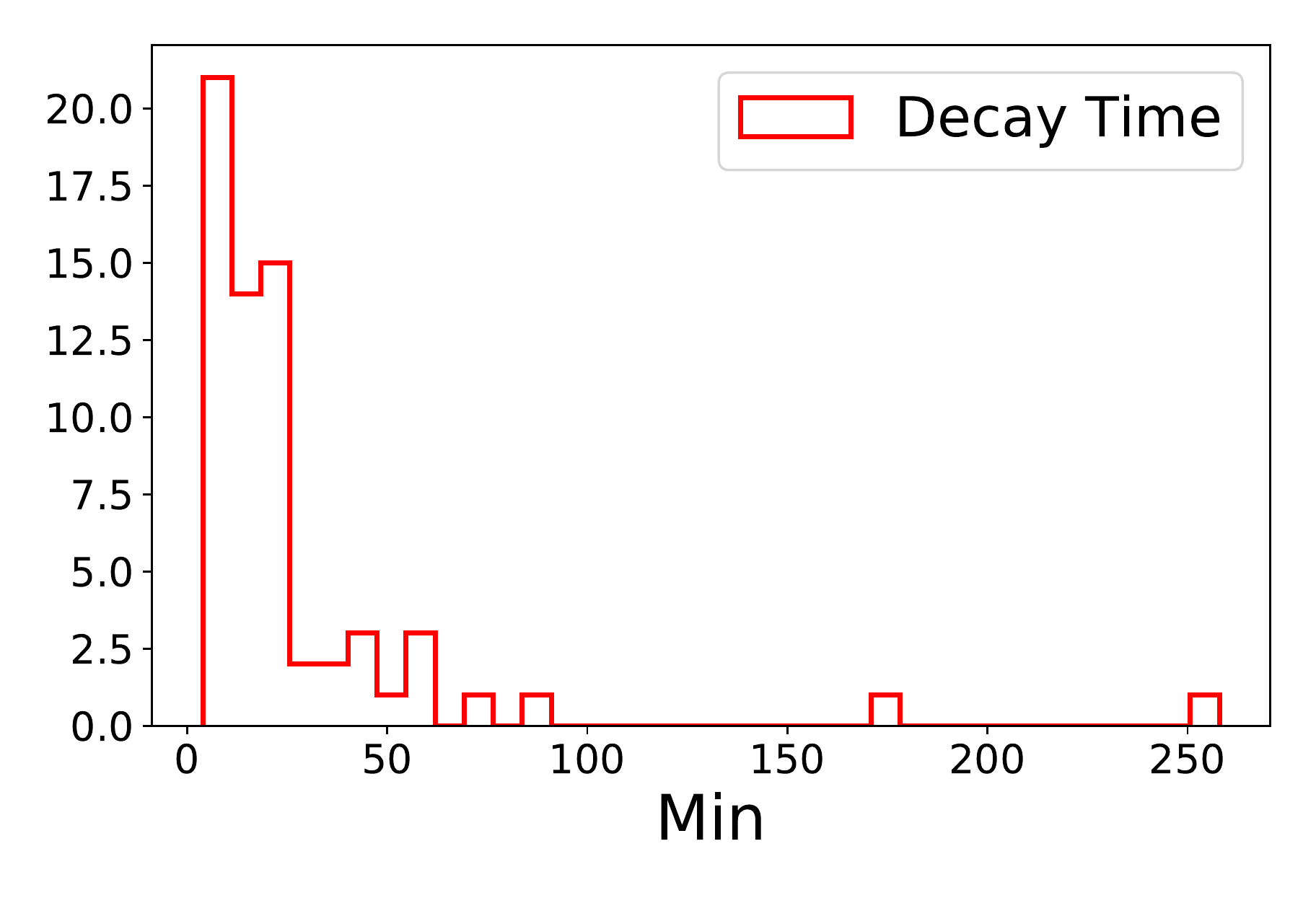}
\includegraphics[scale=0.22]{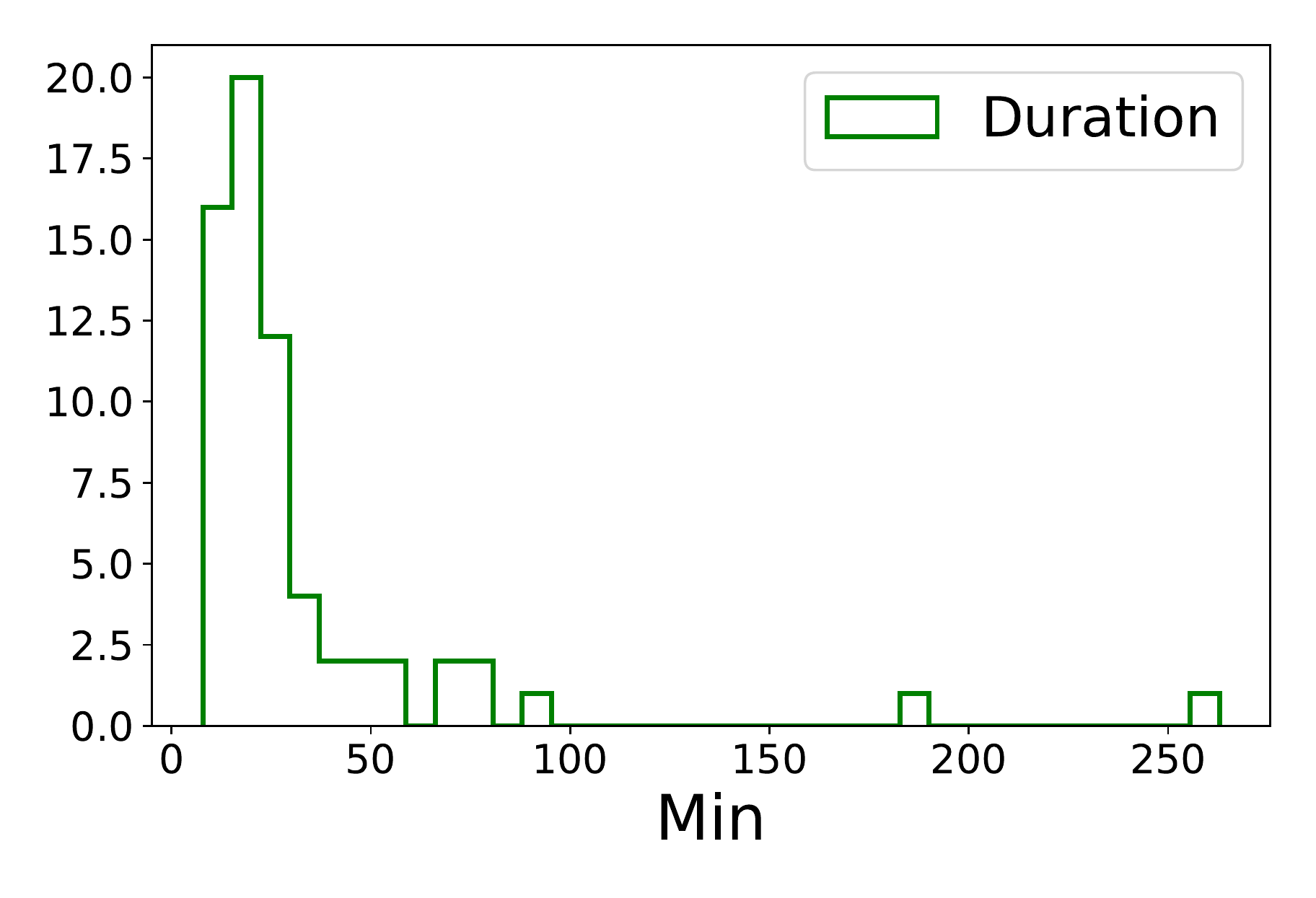}
\includegraphics[scale=0.22]{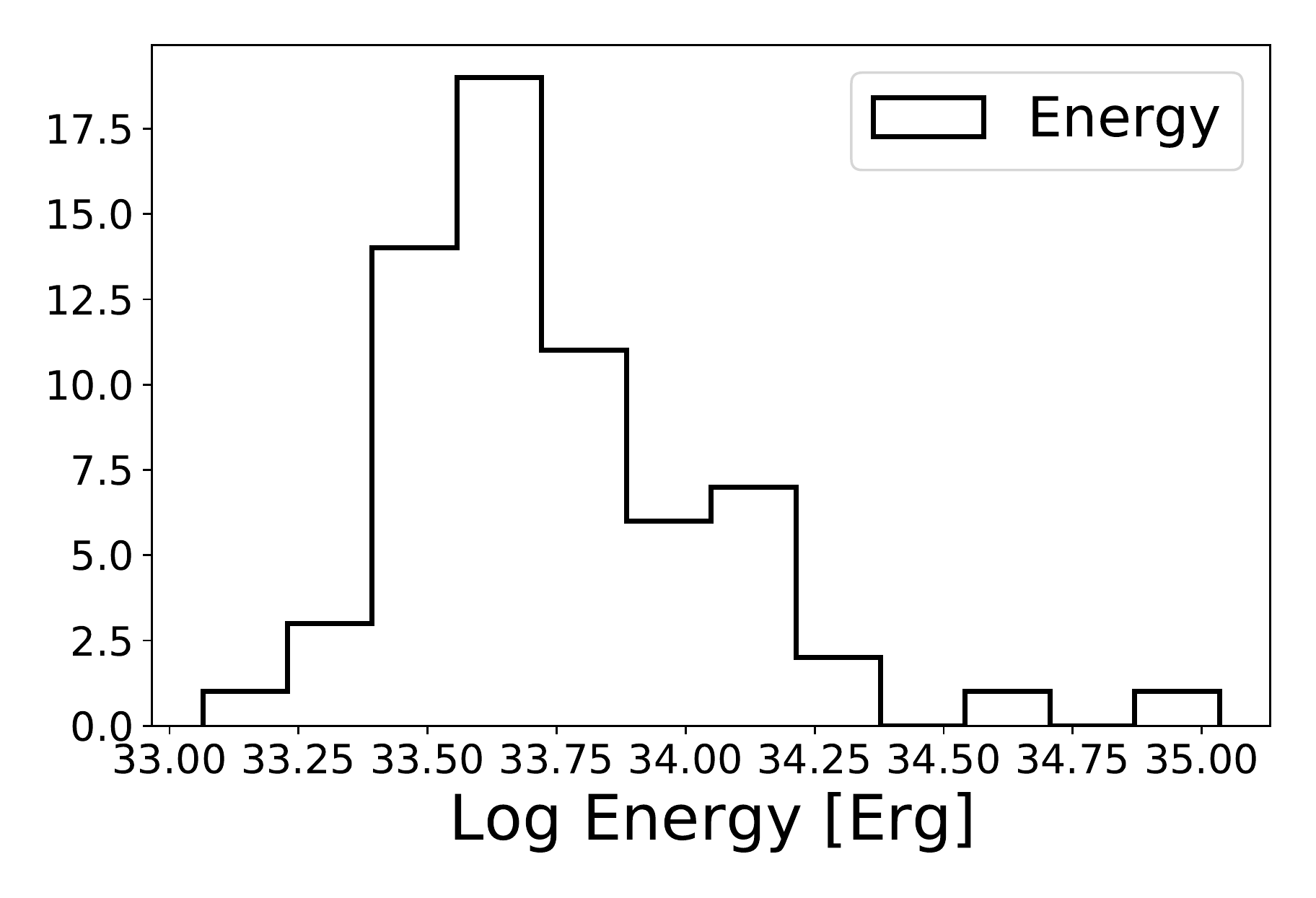} 
\caption{Histograms with statistical results for the flares of Kepler-411:  rise time (top left), decay time (top right), duration (bottom left), and energy (bottom right).}
\label{fig:histflare}
\end{figure}

\begin{figure*}
\centering
\includegraphics[scale=0.4]{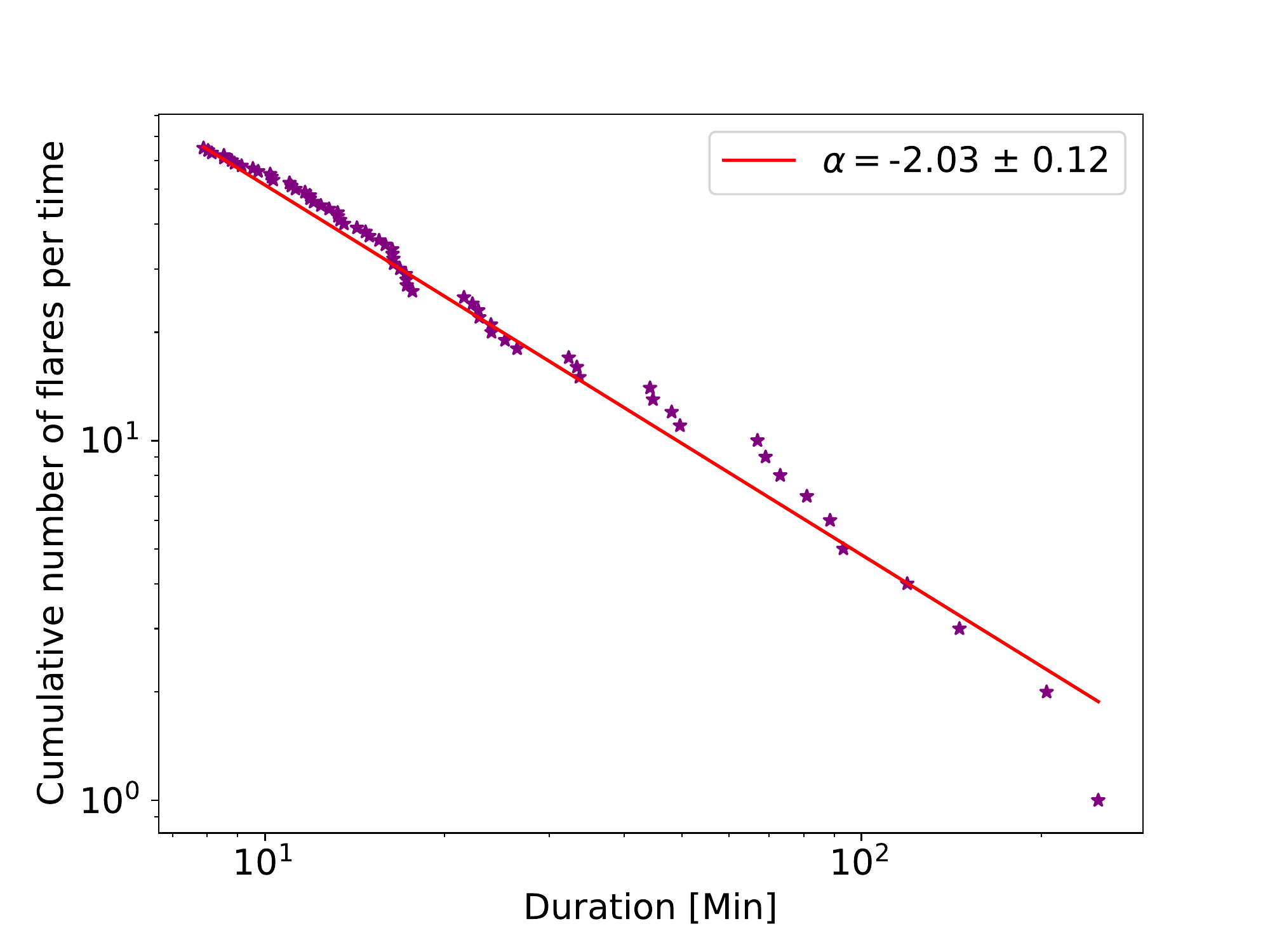} 
\includegraphics[scale=0.4]{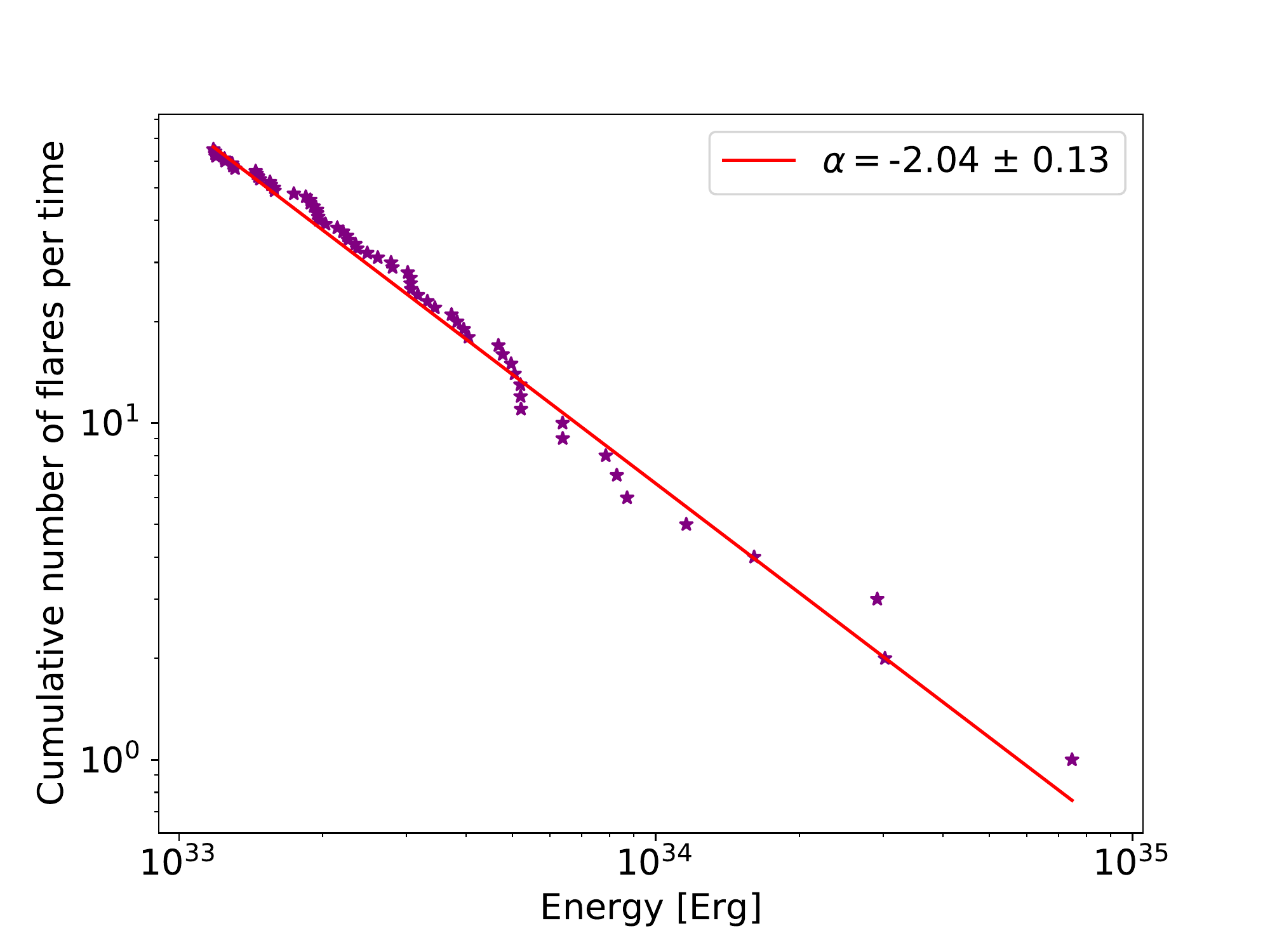} 
\includegraphics[scale=0.4]{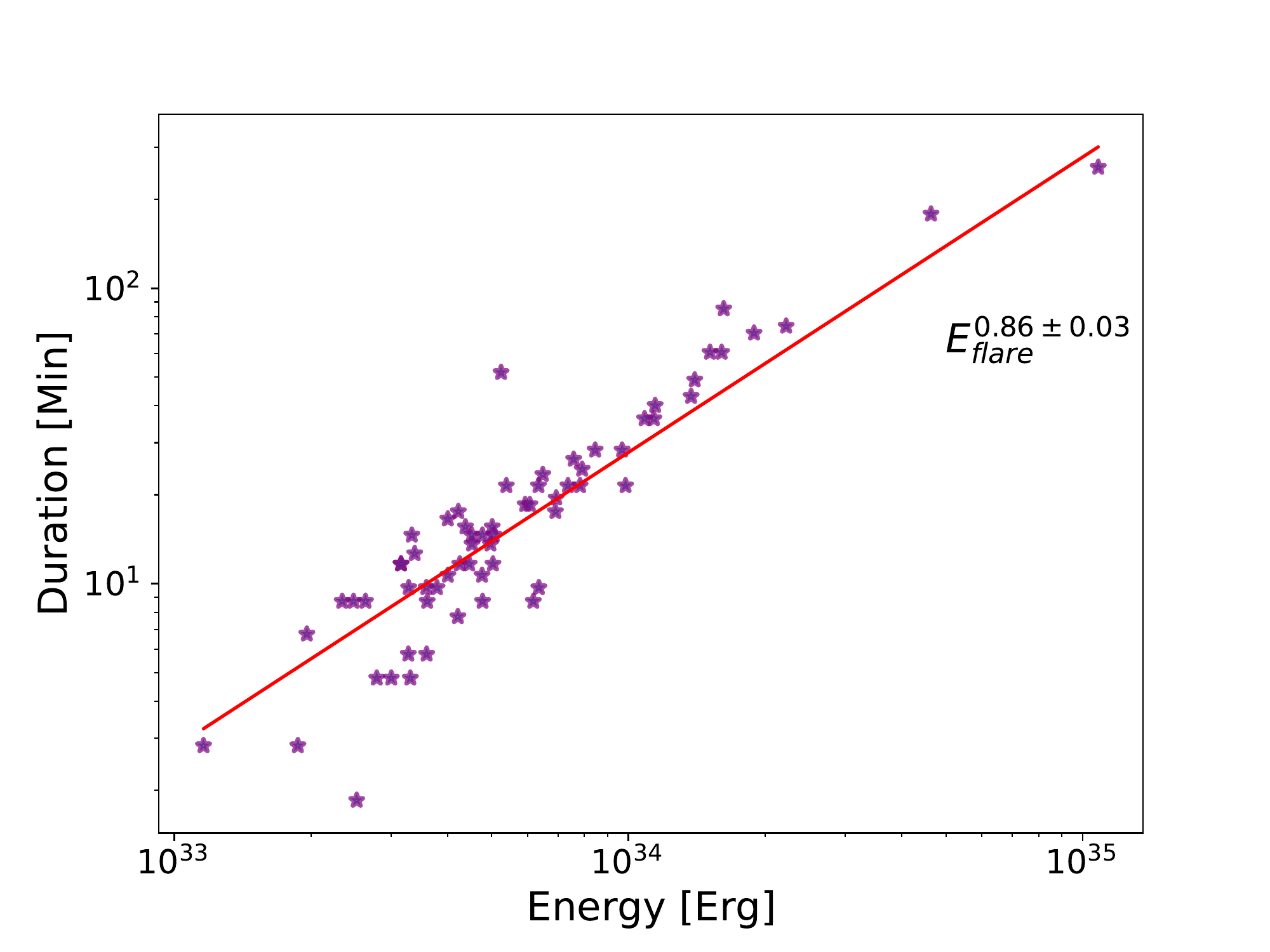} 
\caption{\textbf{Top left:} Fit the power law duration.  \textbf{Top right:} Fit the power law energy. \textbf{Bottom:} Flares duration versus total energy.}
\label{fig:flare+energy}
\end{figure*}

The estimate of the flare energy was performed according to \citet{shiba+13} using the flare luminosity, amplitude, and duration:

\begin{equation}
    E_{flare}=\int L_{flare}(t)dt.
    \label{eq:energy}
\end{equation}

Here we assume that the spectrum of white light flares can be described by that of a blackbody, with an effective temperature of 10,000 K.
Associated errors of stellar parameters such as size and temperature can directly influence  energy measurements of superflares. Moreover, errors in the determination of flare starting and ending points, as well as quiescent levels,  also affect  flare amplitude  and  energy  values by about $\sim$30\% \citep{okamoto2021statistical}.\\

After careful visual inspection, we identified a total of 65 flares on the light curve of the star Kepler-411. 
The energy of each flare was calculated according to Equation~\ref{eq:energy} yielding flare energies in the range from $10^{33}$ to $10^{35}$ erg. For each flare, we also calculated its rise and decay times, as well as total duration in minutes.
Histograms with the  results of the flare properties are presented in Figure~\ref{fig:histflare}. The flares of the K-type star Kepler-411 usually take 5 min to reach their peak flux and then  28 min on average to decay, resulting in a  mean total duration of about 33 min. The statistical results of flares from Kepler-411 are listed in Table~\ref{tab:flare}.\\

The duration of the superflares as a function of their energy is plotted in the bottom  panel of Figure~\ref{fig:flare+energy}, evidencing  a strong correlation (0.88 correlation coefficient) between these properties. 
That is, stronger superflares last  longer. Similar  results  were  also  seen  in  G-type stars as reported by \cite{maehara+15} and by \cite{namekata2017statistical} when studying energy correlations of solar and stellar flares, both of which are believed to be related due to a universal mechanism of magnetic reconnection. Recently, \cite{maehara2021time} investigated flares duration vs flare energy with spectroscopic and photometric observations of the M dwarf flare star YZ Canis Minoris. As a result,  a positive correlation was found  between the duration of the flare, $\tau_{flare}$, and the bolometric energy of the flare, $E_{flare}$ (see Figure 13 of \cite{maehara2021time}), where the duration of the flares increases with energy as $\tau_{flare} \propto$ E$^{0.21 \pm 0.04}_{flare}$.  The authors then compared this result with solar flares for various types of stars. The $E-\tau$ relations show positive correlation, and the power law slope for these $E-\tau$ relations are approximately the same: $\tau_{flare}$ $\propto$ \ $E_{flare}^{1/3}$ \citep{maehara2021time} (see Figure 14 \cite{maehara2021time}).\\

In Kepler 411, we have found the relation  $\tau_{flare}\propto$ E$^{0.86 \pm 0.03}_{flare}$. The  $E-\tau$ value obtained for Kepler-411 is higher than the values obtained by \cite{maehara2021time}. In the literature there are values for the power index for $E$ of, for example, $\sim0.6$  \citep{jackman2021stellar}, 0.42 \citep{tu2021superflares}, and  a diversity of values is presented in Figure 5 of \cite{wu2015}. Probably, the different values are related to the methodology in determining the start and end of flux increase in flares. Here, we use the total duration, where the determination of the start and end times of flares were detailed at the beginning of Section ~\ref{sec:flarecurva}. Another reason may be that the relations refer to stars of different spectral types.\\

Flares are also seen to follow a power--law relation for the energy distribution:
\begin{equation}
    N(E)dE=\beta E^{-\alpha} dE
\end{equation}
\noindent where $\alpha$ and $\beta$ are constants. The power-law fit was performed using a Bayesian Markov chain Monte Carlo \citep{wheatland2004bayesian}, yielding $\alpha$ = 2.04 $\pm$ 0.1 for the Kepler-411 flares energy (seen Figure ~\ref{fig:flare+energy}). This $\alpha$ value is very similar to the results found by \cite{maehara2021time} ($\alpha$ = 2.8 $\pm$ 0.1), \cite{hawlay+14} ($\alpha$ = 1.52 -- 2.32), \cite{lurie2015kepler} ($\alpha$ = 1.92 $\pm$ 0.01 -- 2.03 $\pm$ 0.02), and \cite{wu2015} (2.04 $\pm$ 0.17).\\

The frequency of superflares as a function of their energy and duration is shown in the top panels of  Figure~\ref{fig:flare+energy}.
Predominantly, the superflares have energy of the order of $10^{33}$ erg with superflare frequency given by:
\begin{equation}
  F =  \left(\frac{N_{flares}}{t_{end}-t_{start}}\right).
\end{equation}
The occurrence rate of superflares from  Kepler-411 was estimated to be once every 9 days.

\section{Starspot area versus Superflare energy \label{sec:spotflares}} 

The similarities of the behavior between solar flares and superflares of solar-type stars suggest that solar flares and superflares are caused by the same physical process, \textit{i.e.}, magnetic reconnection \citep{maehara+15}. Therefore, we expect that just like on the Sun, these large flares are associated with the presence of large spots on the stellar surface.\\ 

Finding correlations between stellar superflares and the presence of stellar spots is a challenge for stellar astrophysics, given the technological limitations and the time dedicated by telescopes.
Using the modeling method of starspots with transits, \cite{araujo+21} performed the first mapping of a star surface at three latitudes  using a planetary system. The transit mapping model applied to the three planets yields a total of 198 starspots identified on the surface of the star during the 590 days of observation by the Kepler mission.\\

\begin{figure}[!t]
    \centering
    \includegraphics[scale=0.4]{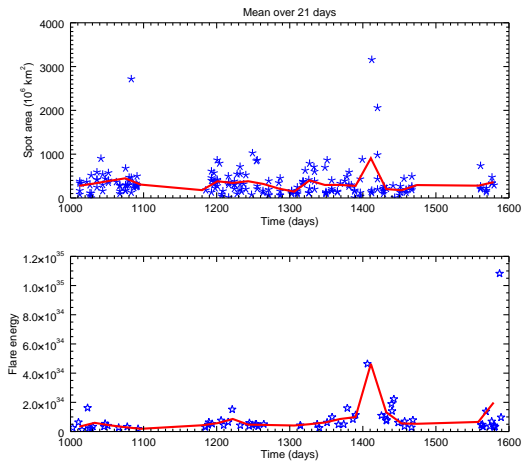}
    \caption{\textbf{Top:}  Starspot areas (blue asterisks), with their running mean every 21 days, overploted as a red line. \textbf{Bottom:} Superflare energy (blue symbols) as a function of time, with the average energy taken every 21 days, overploted as a red line.}
    \label{fig:raio21}
\end{figure}

Then we searched for a correlation pattern of the starspot area with the occurrence of superflares. First, we analyzed the spot area and superflare energy over time looking for coincidences. The area over time of the 198 spots identified in the transit light curves is plotted in the top panel of Figure~\ref{fig:raio21}. An average of the spot area within a time interval of 21 days is plotted as a red curve. The energy of the 65 superflares detected during the same period is plotted in the bottom panel of Figure~\ref{fig:raio21}. Also shown as a red curve is the superflare energy averaged over 21 days. A peak both in spot area as well as in superflare energy is clearly seen around day 1400 BKJD. This suggest an increase in stellar magnetic activity around this period that lasted for about 100 days.\\

\begin{figure}[!htb]
    \centering
    \includegraphics[scale=0.4]{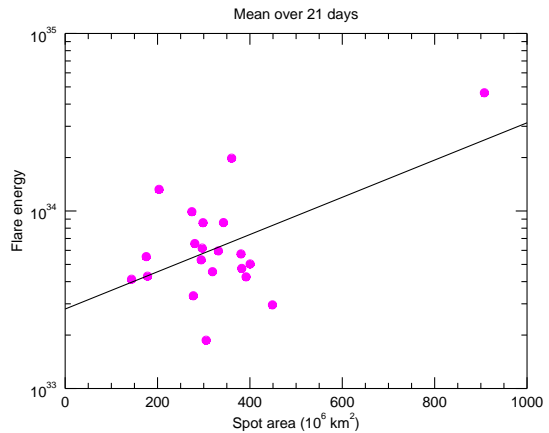}
    \caption{Correlation of Kepler-411 superflare energy with starspot area, averaged every 21 days. }
    \label{fig:energiaxarea}
\end{figure}

To verify the correlation of starspots area and superflare energy suggested in Figure~\ref{fig:raio21}, we plot these averaged quantities against each other in Figure~\ref{fig:energiaxarea}. The solid black line indicates the positive correlation between superflare energy and spot area with a correlation coefficient of 0.54.
In fact, we calculated the average values of spot area and superflare energy for different time intervals varying from 5 to 100 days, and obtained positive correlation only between averages within 16 and 35 days. For other time intervals either there were no correlation or the correlation was negative. The 21 day interval used for the averages yields the higher positive correlation coefficient between spot area and superflare energy. This time interval around 20 days, required for the maximum correlation, could be related to the lifetime of the spots. 

\section{Discussion and conclusions} \label{sec:conclusao}

In this work, we study the magnetic activity of Kepler-411, a K2V type star.  The $\sim 2$\% amplitude brightness variations due to a spotted photosphere and the presence of many superflares qualify it as an active star. Its light curve displays periodic variations compatible with  an average rotation period of 10.5 days, indicating its relative young age.\\

A total of 198 starspots were detected on the surface of Kepler-411 using the planetary transit mapping method, as reported by  \cite{araujo+21}.  
Also 65 flares were identified in the out--of--transit light curve of Kepler-411. Our main goal in this work was to search for a correlation between starspots and  flare occurrence in the Kepler-411 light curve.\\

Our results showed that there is a positive correlation between spot area and superflare energy, similar to what is found in the Sun. However, the correlation depends on the temporal interval used in the averages of both stellar activity indicators. The optimized time window of the correlation between spot area and superflare energy was found to be 21 days, which is probably related to the lifetime of the spots. For temporal averages within periods smaller than 16 days or larger than about 35 days, the correlation is lost.\\

The observed correlation of the 16-35 days, with maximum at 21 days, averages of the spot area versus flare energy  might indicate that statistically within about 2--3 stellar rotations, the active region that flared comes into view. Thus, active regions with lifetime shorter than this would not contribute to the correlation, while if the lifetime of the active regions were much longer, the correlation would be stronger and also occur for longer average periods.
On the Sun, large long duration active regions may last up to 1 maybe 2 solar rotations.\\

Hence, on Kepler-411, a K type star, large stellar spots produced more energetic superflares, consistent with previous results from stars of other spectral types. For example, see Figure 4 of \cite{okamoto2021statistical} where  the upper limit of flare energies tend to increase as the spot area increase. This means that the energy of the superflare is related to the magnetic energy stored within starspots, and the process is probably similar to that of solar flares \citep{okamoto2021statistical}. The correlation of star spot area with the energy of superflares is also observed using the light curve brightness modulation technique \citep{maehara+15}. However, the analysis of starspots and superflares energy by the transit method has a higher spatial precision, despite  being constrained to the stellar latitude band occulted by the exoplanet.
Fortunately, in Kepler-411 there is coverage of three latitude bands, which provides more information about the stellar surface. However, no superflare occurred during a specific transit so as to identify exactly the flaring active region, and thus the spot area.\\

The spots detected by the exoplanet transit method are restricted to only a fraction of the stellar hemisphere. There are certainly many spots unocculted by the planet and consequently  flares that originate from these active regions. This  certainly affect the correlation of spot area and superflare energy. We would like to point out the difficulty in performing such studies, since it is impossible with current observations to determine  which active region produced each superflare. Nevertheless, transit mapping has the advantage of determining the actual spot size and location over the estimates of spot filling factor from out--of--transit light curve \citep{maehara+12}. Thus, given the difficulty of studying these relationships with the available observational quality, our result can be considered quite significant.\\

Statistical studies have shown that stars of type K and M are typically more explosive than other stars \citep{yang2019flare}. The study of stellar spots and  flares is an important ingredient for  stellar dynamo theory, especially for solar-type stars, emphasizing the importance of stellar activity studies. Moreover, new generation of space telescopes are geared towards studying exoplanet atmosphere which are greatly affected by the ``noise" added from stellar activity  \citep{rackham2018transit}.  Since  stellar contamination is not limited to visual wavelengths but may also be very significant in the near-infrared bands, this will likely affect upcoming JWST spectroscopy of transiting exoplanets atmospheres \citep{rackham2018transit}. Thus stellar activity studies, essentially that of starspots, flares, superflares, and magnetic fields, are essential to guide new observations. Some of the most important properties of flares, coronal mass ejections, and energetic particles, including their energies and frequency of occurrence, are determined by the strength of the magnetic field and the size of the stellar active region associated with the starspot area derived from observations \citep{herbst+21}.

\acknowledgments

We are greatful to the anonymous referee for the suggestions that helped improve this work. The authors acknowledge partial financial support from  FAPESP grant \#2013/10559-5 as well as MackPesquisa funding agency. 

\begin{deluxetable}{lllll}
\tablecaption{Parameters of Superflares in Kepler-411}
\tablehead{
\colhead{Duration} & \colhead{Rise Time} & \colhead{Decay time} &
\colhead{Energy} & \colhead{Peak date} \\
\colhead{(min)} & \colhead{(min)} & \colhead{(min)} &
\colhead{(erg)} & \colhead{(BKJD)}
}
\startdata
 14.71185 &4.90381&  9.8080&2.481670e+33&1002.95128 \\
 15.69256 &4.90427& 10.788 &6.349050e+33&1010.87796 \\
  8.82710 &4.90347&  3.9236&1.159830e+33&1015.73762 \\
\enddata
\tablecomments{Table 3 is published in its entirety in the electronic 
edition of the {\it Astrophysical Journal Letters}.  A portion is shown here 
for guidance regarding its form and content.}
\end{deluxetable}
\vspace{5mm}

\bibliography{sample63}{}
\bibliographystyle{aasjournal}



\end{document}